\documentclass[letterpaper, 10 pt, conference]{ieeeconf}  

\usepackage{xcolor}
\usepackage{graphicx}
\usepackage{subcaption}
\usepackage[table]{xcolor}
\usepackage[dvipsnames]{xcolor}
\usepackage{cite}
\usepackage[T1]{fontenc}
\usepackage{tikz}

\newcommand{\qfoil}{%
\tikz[baseline=-0.6ex, scale=0.1]{
  \path[fill=black, domain=0:360, samples=300, smooth]
    plot ({cos(\x)*cos(2*\x)}, {sin(\x)*cos(2*\x)})
    -- cycle;
}}

\IEEEoverridecommandlockouts                              

\title{\LARGE \bf
{A Case Study in Recovery of Drones using Discrete-Event Systems}
}

\author{Liam P. Burns$^{1}$, Dayse M. Cavalcanti$^{2}$, Felipe G. Cabral$^{2}$, Max H. de Queiroz$^{2}$, \\ Melissa Greeff$^{1}$, Publio M. M. Lima$^{2}$, and Karen Rudie$^{1}$ 
\thanks{*This work was supported by Coordenação de Aperfeiçoamento de Pessoal de Nível Superior – Brasil (CAPES) – Finance Code 001, Emerging Leaders in the Americas Program (ELAP) of the Government of Canada, Fundacao de Amparo a Pesquisa e Inovacao do Estado de Santa Catarina (FAPESC) – Edital 21/2024, and the Natural Sciences and Engineering Research Council of Canada (NSERC).} 
\thanks{$^{1}$ The authors are with the Department of Electrical and Computer Engineering and Ingenuity Labs Research Institute,  Queen's University, Kingston, ON, K7L 3N6, Canada 
        {\tt\small 18lpb1@queensu.ca, melissa.greeff@queensu.ca, karen.rudie@queensu.ca}}
\thanks{$^{2}$ The authors are with the Automation and Systems Engineering Program, Federal University of Santa Catarina, Florianópolis, SC, 88.040-900, Brazil 
        {\tt\small dayse.cavalcanti@posgrad.ufsc.br, felipe.gomes.cabral@ufsc.br, max.queiroz@ufsc.br, publio.lima@ufsc.br}}
}

\begin{document}

\maketitle
\thispagestyle{empty}
\pagestyle{empty}

\begin{abstract}
Discrete-event systems and supervisory control theory provide a rigorous framework for specifying correct-by-construction behavior. 
However, their practical application to swarm robotics remains largely underexplored. 
In this paper, we investigate a topological recovery method based on discrete-event-systems within a swarm robotics context. 
We propose a hybrid architecture that combines a high-level discrete event systems supervisor with a low-level continuous controller, allowing lost drones to safely recover from fault or attack events and re-enter a controlled region. 
The method is demonstrated using ten simulated UAVs in the \texttt{py-bullet-drones} framework. 
We show recovery performance across four distinct scenarios, each with varying initial state estimates. 
Additionally, we introduce a secondary recovery supervisor that manages the regrouping process for a drone after it has re-entered the operational region. 
\end{abstract}

\section{INTRODUCTION}

Unmanned aerial vehicles (UAVs) are seeing increasing adoption in a wide range of applications. 
The prevalence of UAVs also creates opportunities to use multiple agents to cooperatively perform complex tasks, such as surveillance, search-and-rescue, and environmental monitoring \cite{Du2025UAVSwarmSurvey}. 
The study of such multi-agent systems is known formally as swarm robotics (SR). 
Systems in SR are vulnerable to a variety of adversarial attacks or communication failures that can compromise mission objectives. 
Consequently, the capacity to recover from such disruptions before mission objectives are compromised is imperative.

Despite extensive work in the literature on swarm coordination \cite{Khaldi2016ICMIC, Jones2018},  the problem of recovery from agent-level failures remains insufficiently addressed. In particular, when individual agents lose sensing capabilities, communication, or localization, mission objectives can quickly become compromised. 
While prior work has considered mechanisms to mitigate the impact of fault events \cite{Parker1998ALLIANCE, Yang2020TII}, most existing approaches focus on maintaining degraded performance or preventing further deterioration, rather than providing a decentralized, formally guaranteed mechanism for returning the swarm to a safe and mission-viable configuration once failures occur.
Consequently, a framework that can enforce safe behavior, and guarantee a return from failure states is essential, yet remains unexplored in the current body of work. 

The literature distinguishes two recovery paradigms for multi-agent systems, \cite{Yang2020TII}: topological recovery (TR) and compositional recovery (CR). 
In TR, recovery is performed on the network communication structure of the swarm \cite{Mou2022JSAC, shi2025fast_k_connectivity_restoration}, while CR focuses on the physical composition and functional roles of the swarm members \cite{Parker1998ALLIANCE,Tahir2023}. 
Most TR methods assume centralized control with global knowledge of agent states (e.g., positions/connectivity) to construct and optimize the swarm’s communication graph. 
This centralization hampers scalability; as team size grows, so does computation and communication time. 
For example the unmodified fast-k connectivity problem formulated as a quadratically constrained program outlined in \cite{shi2025fast_k_connectivity_restoration} introduces $O(n^4)$ binary decision variables, rendering the optimal formulation computationally infeasible beyond very small teams. 
This centralization also introduces a single point of failure at the planner.

Supervisory control theory (SCT) enables the synthesis of correct-by-construction supervisors that enforce safety and ensure nonblocking behavior. 
The application of discrete-event systems (DES) and SCT to SR has  previously been investigated in the literature \cite{Lopes2016,Simon:2023,Ju2020}.
There are several recovery strategies proposed in the literature of DES \cite{Alves:2022, Oliveira:2025, Cavalcanti:2025}.  
However, most of these approaches rely on assumptions that do not hold for SR systems, such as the ability to reset the system to an initial state \cite{Alves:2022} or model a degraded operational region \cite{Oliveira:2025}.

In this work, we propose a new decentralized TR-based recovery strategy for robot swarms focused on the regrouping of agents after a transient failure occurs that separates a member from the group.
To do so, we adapt the recovery strategy proposed in \cite{Cavalcanti:2025}, originally developed for the resynchronization between plant and supervisor after the diagnosis of a fault or attack.
The strategy separates the problem into two layers, a discrete decision layer and a continuous control layer so that we can focus only on the DES recovery strategy.
We show that, if the drone is recoverable, a non-blocking recovery supervisor can be computed to regroup the lost drone to the swarm, avoiding reaching danger or non-signal regions.

\section{PROBLEM FORMULATION}

We consider a swarm of drones, each modeled as a deterministic automaton, operating under the control of a nominal supervisor that enforces mission-level constraints such as safety and coordinated motion.
Each drone is assumed to possess a standard set of sensing, actuation, and communication capabilities commonly found in cooperative UAV systems, including GPS, gyroscope, compass, radio frequency (RF) sensors, and a communication module, which determine the events that can be observed or executed within the DES framework. 
The swarm behaves nominally as long as the drones remain synchronized and within the operational region; however, attacks or faults can cause a desynchronization, rendering the nominal supervisor inadmissible.

In the swarm setting, a desynchronized drone is one whose precise location is no longer known and is therefore represented by a set of possible zones consistent with the events observed during the fault or attack. 
The set of possible zones will be represented by a state estimate in the model. 
In practice, this estimate may include GPS-denied areas, no-fly boundaries, or areas where operations are degraded due to environmental effects, such as strong electromagnetic interference. 
This discrete notion of state estimation contrasts with the continuous-valued estimates typically used in robotics.

The recovery problem is defined as follows: given a drone that has become desynchronized from its nominal closed-loop behavior and an updated post-desynchronization state estimate, synthesize a recovery supervisor that $(i)$ guarantees the avoidance of unsafe states, $(ii)$ ensures convergence back to a defined operational region, and $(iii)$ enables the drone to rejoin the swarm and resume coordinated operation.
We restrict our analysis to scenarios in which at most one drone becomes lost at any given time. 
While this assumption poses no theoretical limitation within SCT framework, further work is needed to address collision-free behavior when multiple drones undergo recovery simultaneously at the continuous level.

\subsection{System Description}

\begin{figure}
    \vspace{7pt}
    \centering
    \includegraphics[width=0.7\linewidth]{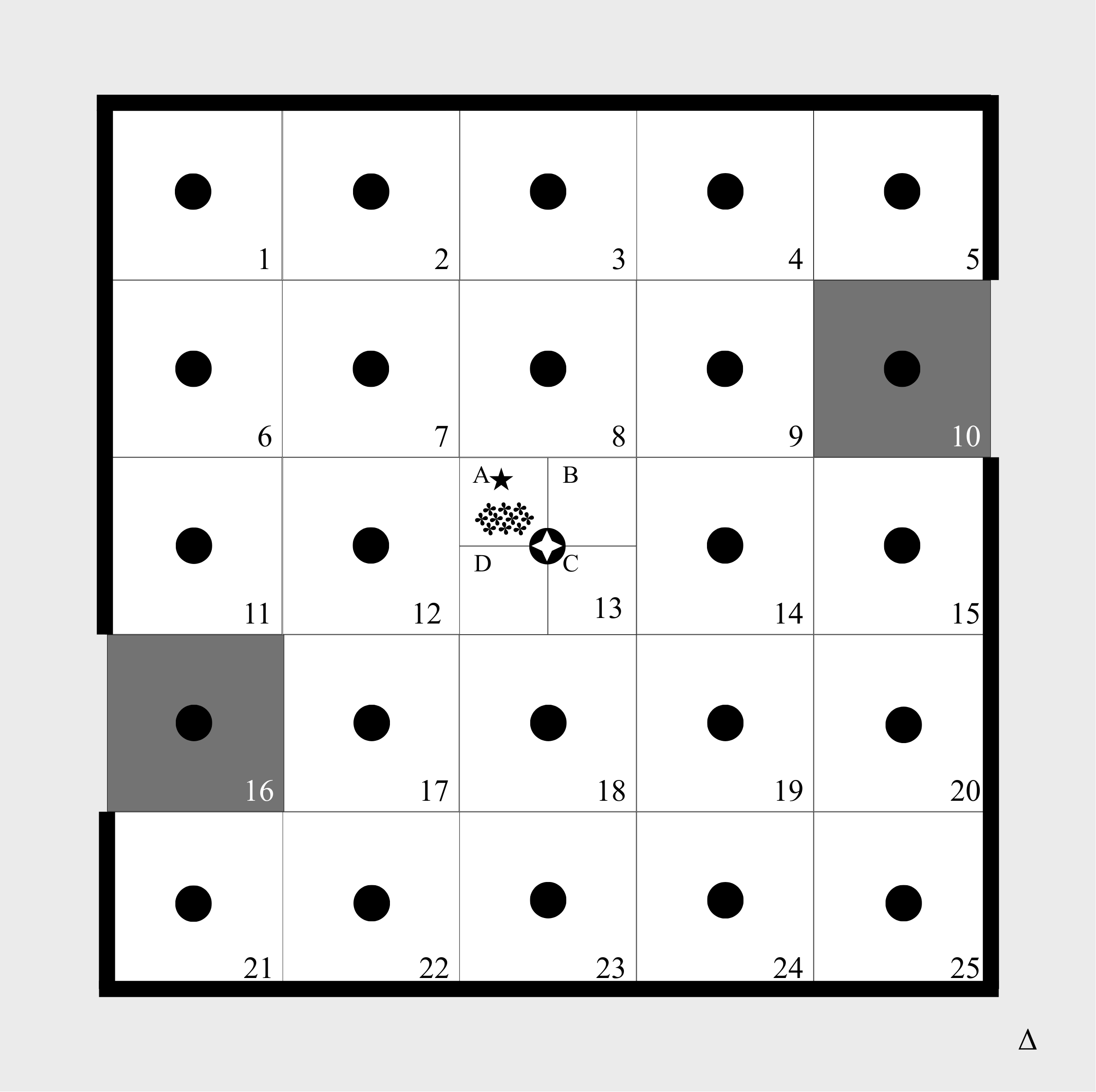}
    \caption{Mission area grid map. Zones OR, BZ, and NFZ are represented by cell $13$, $1-25$ except $13$, and $\Delta$, respectively. Zone $13$ is divided into sub-zones $A-B$. Zones $10$ and $16$ (dark gray) represent unsafe areas where navigation increases the risk of drifting into NFZ. The black star represents the base, the black quatrefoils represent the drone swarm and the black circles represent the reference markers.}
    \label{fig:map1}
    \vspace*{-5mm}
\end{figure}

Consider the grid map shown in Figure~\ref{fig:map1}.
A drone (\qfoil) swarm operates collaboratively to fulfill a mission, coordinating its movements to effectively patrol a designated area.
We distinguish three regions that reflect both mission objectives and operational constraints: the Operational Region (OR, zone $13$), the Buffer-Zone (BZ, zones $1-25$, excluding zone $13$), and the No-Fly Zone (NFZ, zone $\Delta$).
In OR, navigation is reliable, and the swarm operates under nominal conditions.
Precise GPS readings allow each drone to determine its exact position within the sub-zones ($A - D$), with sub-zone $A$ locating the swarm's base of operations ($\star$).
The BZ represents an area where the drones cannot determine their exact location; however, they can still rely on the RF sensors to determine the position of the reference markers ($\bullet$) in the center of the zones to infer movement between the zones. 
The NFZ region corresponds to a outer-perimeter area that must be strictly avoided.
Additionally, zones $10$ and $16$, represented as colored in gray in Figure \ref{fig:map1}, are considered to be unsafe for navigation due to threats or interferences that can cause drones to drift into NFZ without detection.
Moreover, the map is adaptable and scalable as both the base and unsafe regions can be relocated or redefined to represent different mission scenarios.

\begin{figure}
    \vspace{7pt}
    \centering
    \includegraphics[width=0.5\linewidth]{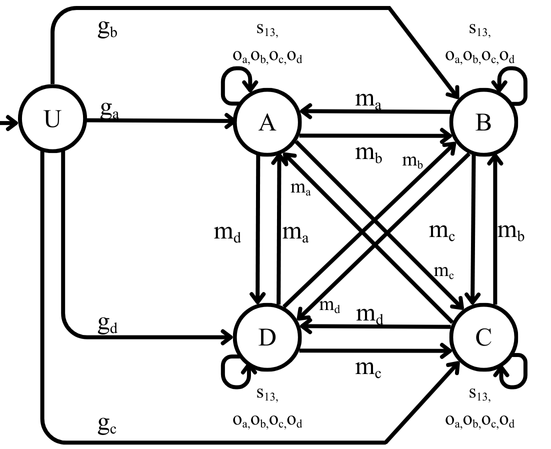}
    \caption{Model of the inner navigation layer in zone $13$,~$G_O$. States $A-D$ model the sub-zones, while $U$ is an unknown state. The events indicate the signals ($g_i$, $i=\{a,b,c,d\}$), permissible movements ($m_i$) between adjacent sub-zones after the search ($s_{13}$), and observation of a border ($o_i$).}
    \label{fig:G4}
    \vspace*{-5mm}
\end{figure}

The drone swarm patrols the OR following a predetermined cyclic route ($A \rightarrow B \rightarrow C \rightarrow D \rightarrow$ repeat) under the control of a nominal supervisor, ensuring organized coverage of the search area.
Figure~\ref{fig:G4} depicts the automaton that models the behavior of the inner navigation layer of zone $13$.
Each state in $G_{O}$ corresponds either to an unknown state ($U$) or to one of the sub-zones ($A-D$), and transitions represent the signals or possible movements between adjacent sub-zones.
The initial state, $U$, waits for the triggering of an event that identifies which sub-zone the drone is currently located at, only then does the automaton transition to one of the states $A$, $B$, $C$, or $D$.
The other automaton for this layer is omitted for brevity.
It merges the local exploration behavior with the search procedure ($search \rightarrow observe \rightarrow move$).
Thus, the drone can reach any neighboring sub-zone from its current location, and after each movement (or return) it may freely select a new scanning direction. 

\begin{figure}
    \vspace{7pt}
    \centering
    \includegraphics[width=0.9\linewidth]{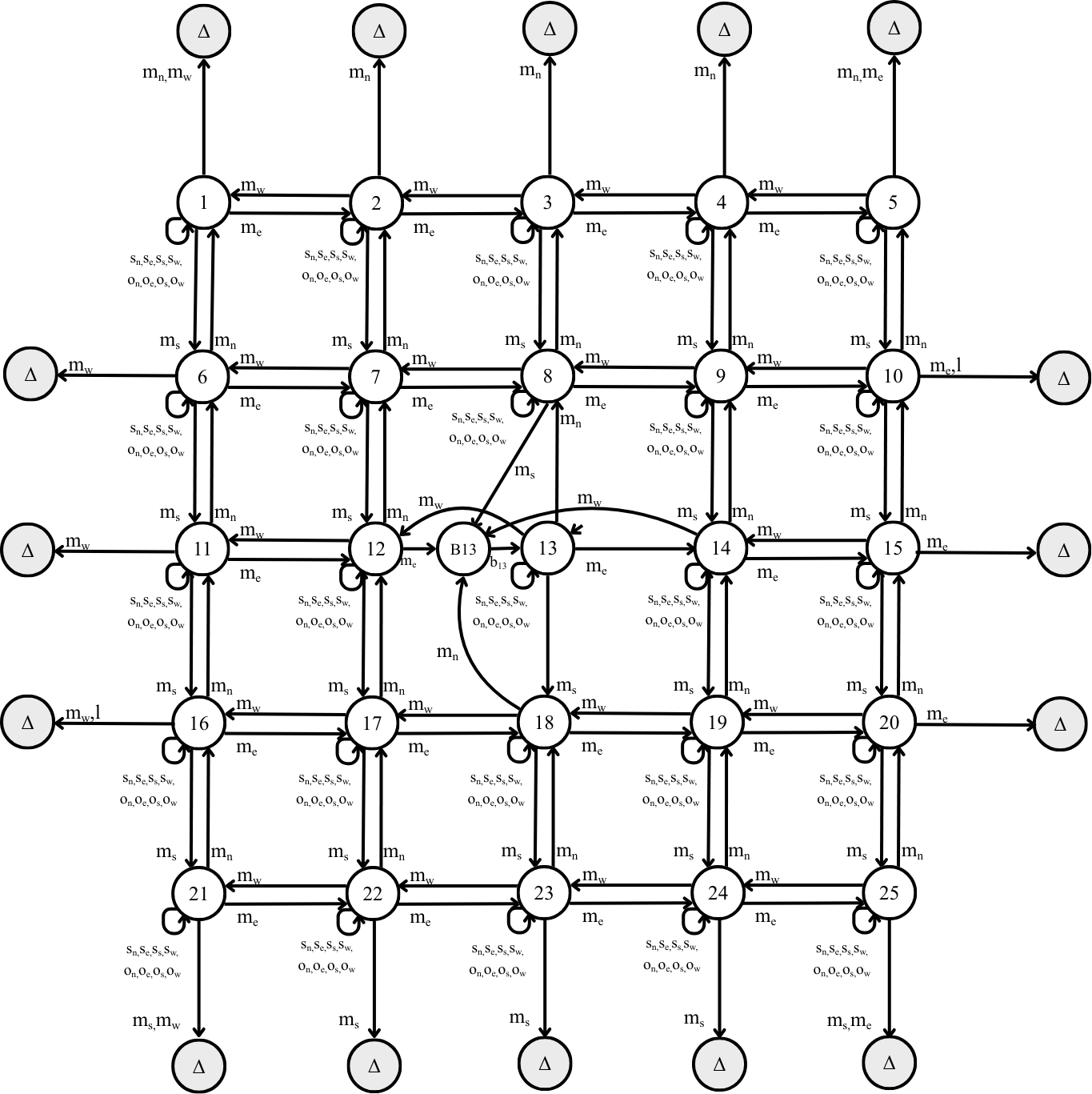}
    \caption{Automaton $G_M$, modeling the navigation layer. Each state corresponds to a zone ($1-25$, $\Delta$), with transitions indicating possible movements ($m_i$, $i \in \{n,e,s,w\}$) between adjacent zones after the search ($s_i$) and observation of a border ($o_i$) or a loss ($l$). The OR ($13$) is modeled as a single state, and an intermediate state ($B13$) is added between $13$ and its neighboring zones to represent detection of re-entry into the OR ($b_{13}$). State $\Delta$ is unsafe.}
    \label{fig:G1}
    \vspace*{-8mm}
\end{figure}

Figure~\ref{fig:G1} depicts the automaton that models the behavior of the OR navigation layer, which describes where the drone can move and the spatial constraints of the environment.
Each state in $G_{M}$ corresponds to either a zone from the BZ ($1-25$, excluding zone $13$), zone $13$ ($13$), the border of zone $13$ ($B13$) or the NFZ ($\Delta$), and transitions represent the possible movements between adjacent zones.
State $13$ is the original initial state and corresponds to OR in the normal operation.
In $G_M$, $13$ is treated as a single state, even though it is subdivided into $A$, $B$, $C$, and $D$ in the other control layer. 
To model the moment at which a drone detects that it has entered the operational region, we include state $B13$.
State $\Delta$ is an unsafe state, but states $10$ and $16$ have a transition with an uncontrollable and unobservable event $l$ (\textit{lost}) to $\Delta$.
Thus, $G_M$ represents the physical layout of the environment, the grid map and the connections between zones.
The size of this automaton grows with the number of zones to be explored; consequently it is natural to ask whether this layer could be represented by a collection of smaller automata.
However, even in this case, the overall model would still exhibit growth in complexity.
For our purposes, we opted for maintaining it as a single automaton to provide a more comprehensive view of the spatial connections and constraints.
The other automata for this layer are ommited for brevity.
In summary, one automaton models the local exploration cycle within a zone, with states representing the stages of \textit{roaming} a zone, \textit{observing} zone boundaries, and awaiting a \textit{movement} control decision ($m$ for \textit{move to the next zone} or $r$ for \textit{return to the current zone}).
A second automaton models the directional scanning behavior, enforcing the required sequence of border checks (\textit{north} $\rightarrow$ \textit{east} $\rightarrow$ \textit{south} $\rightarrow$ \textit{west}) each time the drone enters a new zone.
The explicit search logic was introduced to prevent the drone from becoming trapped in repetitive exploration patterns, detecting and returning to the same border without making progress towards recovery.



\subsection{Recovery Strategy}

\begin{figure*}
    \vspace{7pt}
    \centering
    \includegraphics[width=1\textwidth]{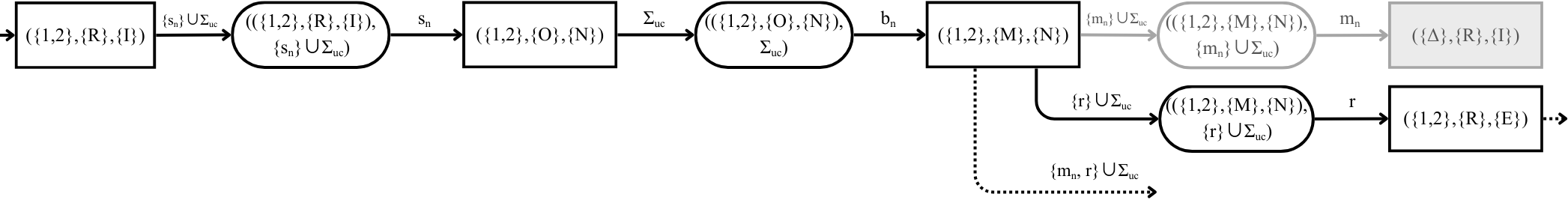}
    \caption{(Part of the) Recovery Bipartite Transition System $T$ from the post-desynchronization state estimate $(\{1,2\},\{R\},\{I\})$, where $\{1,2\}$ is $G_M$ post-desynchronization state estimate and $\{R\}$ (\textit{roaming}) and $\{I\}$ (\textit{idle}) are the post-desynchronization state estimates of the other automata related to the OR navigation layer.}
    \label{fig:T}
    \vspace*{-6mm}
\end{figure*}

Our DES recovery will include two phases. 
In the first phase we develop a primary recovery supervisor to bring a lost drone back into the operational region and in the second phase we develop a secondary supervisor to reconnect that drone with the remaining swarm. 
For the primary supervisor, we employ the notion of recoverability introduced in \cite{Cavalcanti:2025} to assess whether the system, given a post-desynchronization state estimate, can be driven back to its nominal state estimate while avoiding unsafe states. 
To this end, \cite{Cavalcanti:2025} proposes an algorithm for synthesizing a Recovery Bipartite Transition System (RBTS) and, subsequently, a Recovery Supervisor.
The RBTS is a game-structure that models the interaction between a supervisor and a plant, where each one plays based on the plant state estimate and the feasible event set.
The RBTS has two types of states: $Y$-states and $Z$-states.
The $Y$-states (represented by rectangles in Figure~\ref{fig:T}) correspond to the supervisor's turn, $i.e.$, 
states where the supervisor must decide the set of events to be enable in the plant including all uncontrollable events of $\Sigma_{uc}$ and a subset of the eligible controllable events at the current state estimate.
Thus each Y-state is associated to a state estimate represented as a tuple of subsets of states of the automata in the composite plant, and each outgoing transition is associated to a control decision enabling at least one feasible event in the plant.
The $Z$-states (represented by obrounds in Figure~\ref{fig:T}) correspond to the plant's turn, which are the states where the plant can generate a feasible observable event based on the previous control decision.
They are formed by the current state estimate of the plant and the last enabled control decision.
In this scheme, each transition of the RBTS is from a $Y$-state to a $Z$-state or vice-versa, depending on whose turn it is.
The transitions from a $Y$-state to a $Z$-state occur with the chosen control decisions, while the transitions from a $Z$-state to a $Y$-state occur with the observation of an event.
That way, each branch of the RBTS corresponds to a distinct sequence of control decisions and observations.
The game is initiated by considering that the first player is the supervisor and, thus, the initial state of the RBTS is a $Y$-state formed by the unobservable reach of the initial state of the plant.
By the end of the computation, the RBTS finds a possible and safe recovery path from an uncertain state estimate to a nominal state estimate or returns an empty BTS if the system is not recoverable.
A recovery supervisor is therefore a winning strategy extracted from the RBTS, selecting, at each state, a control decision that guarantees the recoverability objective against all admissible plant behaviors.

Figure~\ref{fig:T} depicts the initial steps in the construction of the RBTS starting from the post-desynchronization state estimate $(\{1,2\},\{R\},\{I\})$, where $\{1,2\}$ corresponds to the state estimate of $G_M$ post-desynchronization and $\{R\}$ (\textit{roaming}) and $\{I\}$ (\textit{idle}) are the post-desynchronization state estimates of the other automata related to the OR navigation layer.
In this $Y$-state, the control action $\{s_n\}\cup \Sigma_{uc}$ (where $s_n$ is \textit{search north}) is the only one available for all automata.
By applying this control decision, the system transitions to $Z$-state $((\{1,2\},\{R\},\{I\}),\{s_n\} \cup \Sigma_{uc})$.
Upon the observation of event $s_n$, the system transitions to $Y$-state $(\{1,2\},\{O\},\{N\})$, where $G_M$ remains in state estimate $\{1,2\}$, while the other automata advance to states $O$ (\textit{observing zone boundary}) and $N$ (\textit{searching for the northern border}).
At this point, no other controllable events can be enabled and the only available control decision consists solely of uncontrollable events.
Upon the observation of event $b_n$ (\textit{northern border has been found}), one of the automata advances to state $M$ (\textit{awaiting a movement control decision}).
From this configuration, three control decisions are admissible: $\{m_n\}\cup \Sigma_{uc}$ (where $m_n$ is \textit{move north}), $\{r\}\cup \Sigma_{uc}$ (where $r$ is \textit{return}), and $\{m_n, r\}\cup \Sigma_{uc}$.
The branch for control decision $\{m_n\}\cup \Sigma_{uc}$ is eliminated from the RBTS (shaded in Figure~\ref{fig:T}) as it would lead $G_M$ to unsafe state $\Delta$, thus, the following control decision, $\{r\}\cup \Sigma_{uc}$, is chosen.
Note that the branch for control decision $\{m_n,r\}\cup \Sigma_{uc}$ is not explored since a winning strategy is already obtained by selecting $\{r\}\cup \Sigma_{uc}$.
Due to space limitations, we omit the detailed computation of the recovery supervisor. 
Readers may refer to \cite{Cavalcanti:2025} for the full algorithmic procedure.

However, the recovery process is not complete once the drone reenters the OR.
To fully restore the system, the drone must be regrouped with the swarm and resynchronized with its nominal supervisor.
Simply resuming the cyclic route ($A \rightarrow B \rightarrow C \rightarrow D \rightarrow$ repeat) before regrouping may cause the drone to miss the swarm's current position as it continues to patrol the zone. 
To address this, we construct a secondary supervisor that takes care of reconnecting the drone with the swarm.
Two natural regrouping strategies can be considered: 1) the drone may remain in a sub-zone within the operational region and wait for the swarm to pass by, or 2) it may patrol in the reverse order ($D \rightarrow C \rightarrow B \rightarrow A \rightarrow$ repeat) to increase the likelihood of the encounter.
We use the first strategy.
We represent the supervisor that implements this strategy with a two-state finite-state machine (not shown due to space limitations) that has events alternating between searching for the swarm ($s_{13}$) and returning ($r$).

In \cite{Cavalcanti:2025}, a recovery structure is proposed to handle the switching between supervisors. 
However, this structure is not designed to operate as a supervisor itself, but rather as part of a higher-level architecture. 
In this work, we propose a finite-state machine in which events explicitly trigger the switching mechanism. 
Under this formulation, the overall system is nonblocking, provided that the nominal supervisor is nonblocking and the system is recoverable.

\section{IMPLEMENTATION AND SIMULATION FRAMEWORK}

\subsection{Hybrid System Architecture}

In order to simulate the system, the overall control architecture is divided into two layers: a discrete decision layer and a continuous control layer.
The discrete layer is modeled as a DES and synthesized using SCT, handling high-level logic such as mode selection (e.g., nominal or recovery) and determining the active grid cell.
The continuous layer is implemented using a low-level controller which governs the drone’s dynamics and translates high-level discrete commands into feasible control inputs.
In this study, we use a model predictive controller (MPC) \cite{Andersson2019} for this purpose.

This hierarchical decomposition cleanly separates what the system is allowed to do from how it executes those instructions. 
This configuration has two main advantages. 
First, it improves modularity, i.e., the separation of the DES layer and the continuous layer allows each component to be swapped with minimal modification to the other.
Second, it allows for the direct use of SCT synthesis results, ensuring that supervisors are non-blocking, maximally permissive, and controllable. 
This is particularly beneficial during verification processes. 
Non-SCT approaches need to manually verify numerous corner cases to ensure correct behavior, while SCT-based controllers are correct-by-construction.

\subsection{Supervisory Control Translated To  Microcontrollers }

In classical SCT, the plant $G$ is fixed and the designer cannot alter the events it generates. 
In microcontroller-based systems, by contrast, the event generation is largely implemented in software. 
Although the system may react to external events (e.g., sensor inputs or interrupts), how it responds is entirely user-defined. 
As a result there is no clear analogue of an independent plant $G$ separate from the control logic.   
To preserve the plant–supervisor separation required by SCT, we explicitly partition the system logic into two components. The first component is the plant, specifying the low-level behavior executed while the system remains in a given state or grid cell (e.g., the order in which the drone moves). 
The second component is the supervisor, which dictates when transitions to new states or behaviors are allowed to occur, enabling movement to new zones based on the current state and observed events.
This setup allows us to maintain the conceptual separation of plant and supervisor assumed in SCT, while implementing both within the same programmable microcontroller architecture.




\subsection{Experimental Setup}

We first simulate the system in \texttt{Supremica}.
Then, we evaluate the proposed architecture in the \texttt{gym-pybullet-drones} environment \cite{panerati2021learning}. 
Ten drones are created with initial positions evenly distributed across sub-zone $A$ of OR. 
Each drone maintains local copies of the nominal supervisor and both the primary and secondary recovery supervisors in memory. 
At time $t=0$, one of the ten drones is selected at random to experience a fault event, after which it is relocated to a zone within the BZ area. 
From here, the primary recovery supervisor is (re)calculated based on an initial estimate of where the drone has landed, using the algorithm outlined in \cite{Cavalcanti:2025}.
The lost drone subsequently executes its recovery search behavior, transitioning into new cells only when enabled by the recovery supervisor. 
Meanwhile, the remaining drones in the swarm execute the nominal behavior.
Eventually, when the lost drone returns to the nominal region, the drone switches to the secondary recovery supervisor, shown in Figure \ref{fig:SimS3}, where it waits until it is sufficiently close to the swarm, before resuming nominal operations. 
The simulator executes the control loop in discrete time, with a control period of $\Delta t = \frac{1}{240}s$, where 240s is the default simulation period in \texttt{gym-pybullet-drones}.
Thus, the simulation time is defined as $t = k \Delta t$ where $k$ is the iteration index of the main control loop.


\section{RESULTS}


To evaluate the effectiveness of the proposed architecture, four trials are conducted, each designed to assess system performance under different recovery scenarios. 
The primary recovery supervisors are obtained directly from the algorithm proposed in \cite{Cavalcanti:2025}.
Each trial begins with an initial state estimate, defined as the set of cells where the recovery system believes the lost drone may be. 
In each trial, the drone is initialized in one of the cells of the initial state estimate and then guided back to nominal control by the primary recovery supervisor. 
The recovery time, defined as the time from a drone’s departure from the initial state in the uncontrolled area to its re-entry into the OR, is measured in simulation and recorded in all trials.

Table~\ref{tab:recovery_times} summarizes the result, however, it is important to note that the recovery time metric presented is used only as a way of demonstrating the impact of the different initial state estimates on the RBTS, and is not meant to demonstrate optimality of any given path.
The algorithm proposed in \cite{Cavalcanti:2025} successfully synthesized primary recovery supervisors for three of the four trials. In the fourth trial, since the state estimate corresponds to the entire top region ($1 -5$) of the map, the system is not recoverable because there are no events the supervisor can enable that would guarantee avoidance of the unsafe states.
In the simulation, this manifests as stalling behavior, since the supervisor cannot safely authorize any movement.
We can also observe that, despite the fact that all other trials included zone $1$ in the initial state estimate, their recovery times  differ due to the additional states present in each estimate.
These variations highlight how the composition of the state estimate influences both the availability of safe actions and the efficiency of the recovery process.
When the estimate is large or highly uncertain, the recovery supervisor must consider a broader set of safe actions and successor estimates, increasing the branching factor of the RBTS and, consequently, the computational effort required to identify a valid recovery path.

\begin{table}[t]
    \centering
    \caption{Recovery performance across simulation trials. Each trial corresponds to a unique post-desynchronization state estimate and starting cell.}
    \label{tab:recovery_times}
    \resizebox{\columnwidth}{!}{%
    \begin{tabular}{|c|c|c|c|c|}
        \hline
        Trial & Initial Estimate & Starting Cell & Recovery Time (s)\\ \hline

        1 & \{1, 2\} & 1 & 38.1 \\ \hline

        1 & \{1, 2\} & 2 & 32.5 \\ \hline

        2 & \{1, 6, 11\} & 1 & 110.3 \\ \hline

        2 & \{1, 6, 11\} & 6 & 129.0 \\ \hline

        2 & \{1, 6, 11\} & 11 & 16.4 \\ \hline


        3 & \{1, 2, 6, 7\} & 1 & 65.9 \\ \hline

        3 & \{1, 2, 6, 7\} & 2 & 262.5 \\ \hline
 
        3 & \{1, 2, 6, 7\} & 6 & 27.5  \\ \hline

        3 & \{1, 2, 6, 7\} & 7 & 122.9  \\ \hline

     
        4 & \{1, 2, 3, 4, 5\} & 1/2/3/4/5 & No Solution Found \\ \hline

    \end{tabular}%
    }
    \vspace*{-5mm}
\end{table}

\begin{figure}[h]
    \vspace{7pt}
    \centering
    \begin{subfigure}{0.23\textwidth}
        \includegraphics[width=\textwidth]{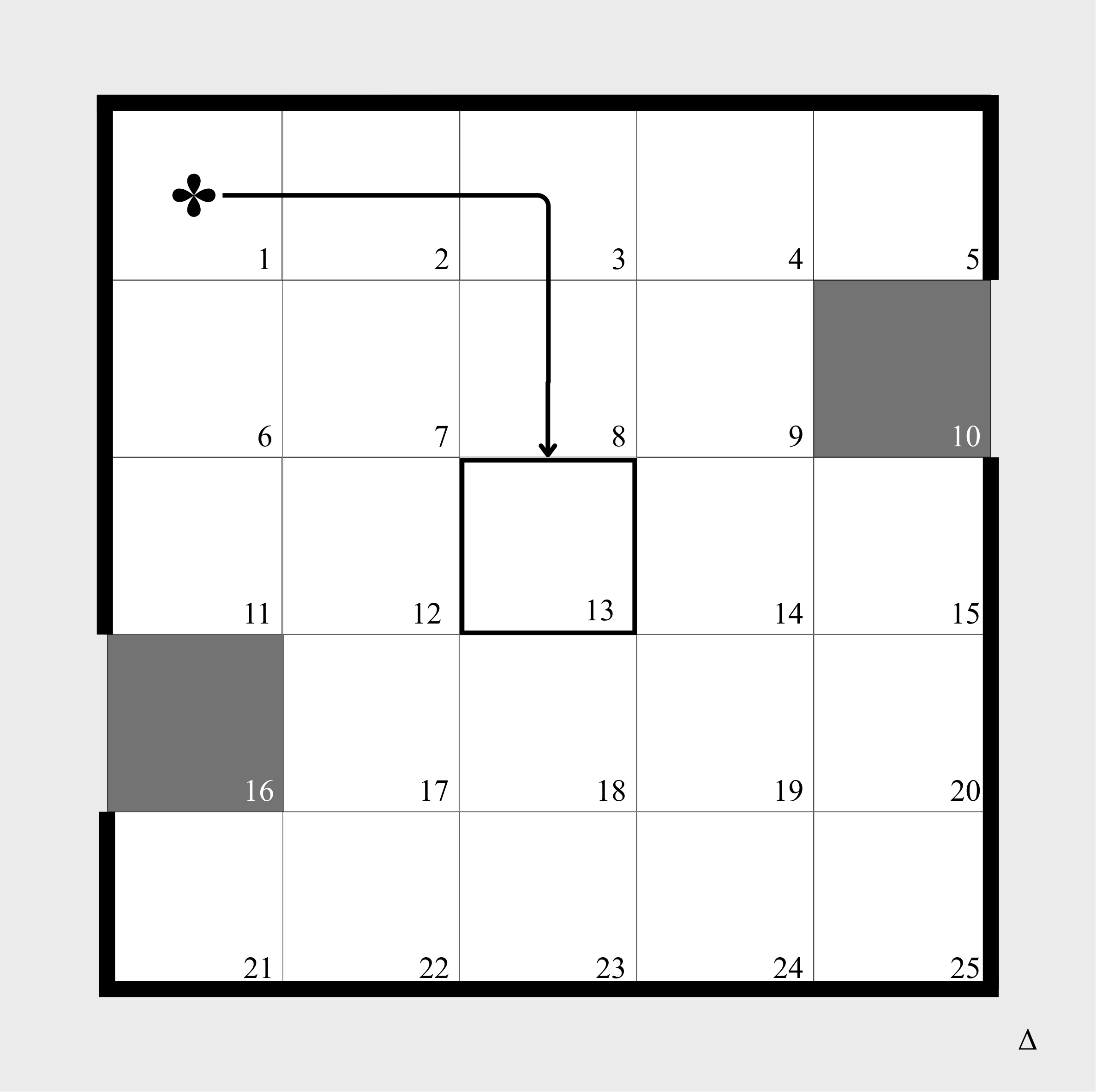}
        \caption{Drone in zone $1$ and initial state estimate = $\{1,2\}$.} 
        \label{fig:S1Z12D1}
    \end{subfigure} 
    \hfill
    \begin{subfigure}{0.23\textwidth}
        \includegraphics[width=\textwidth]{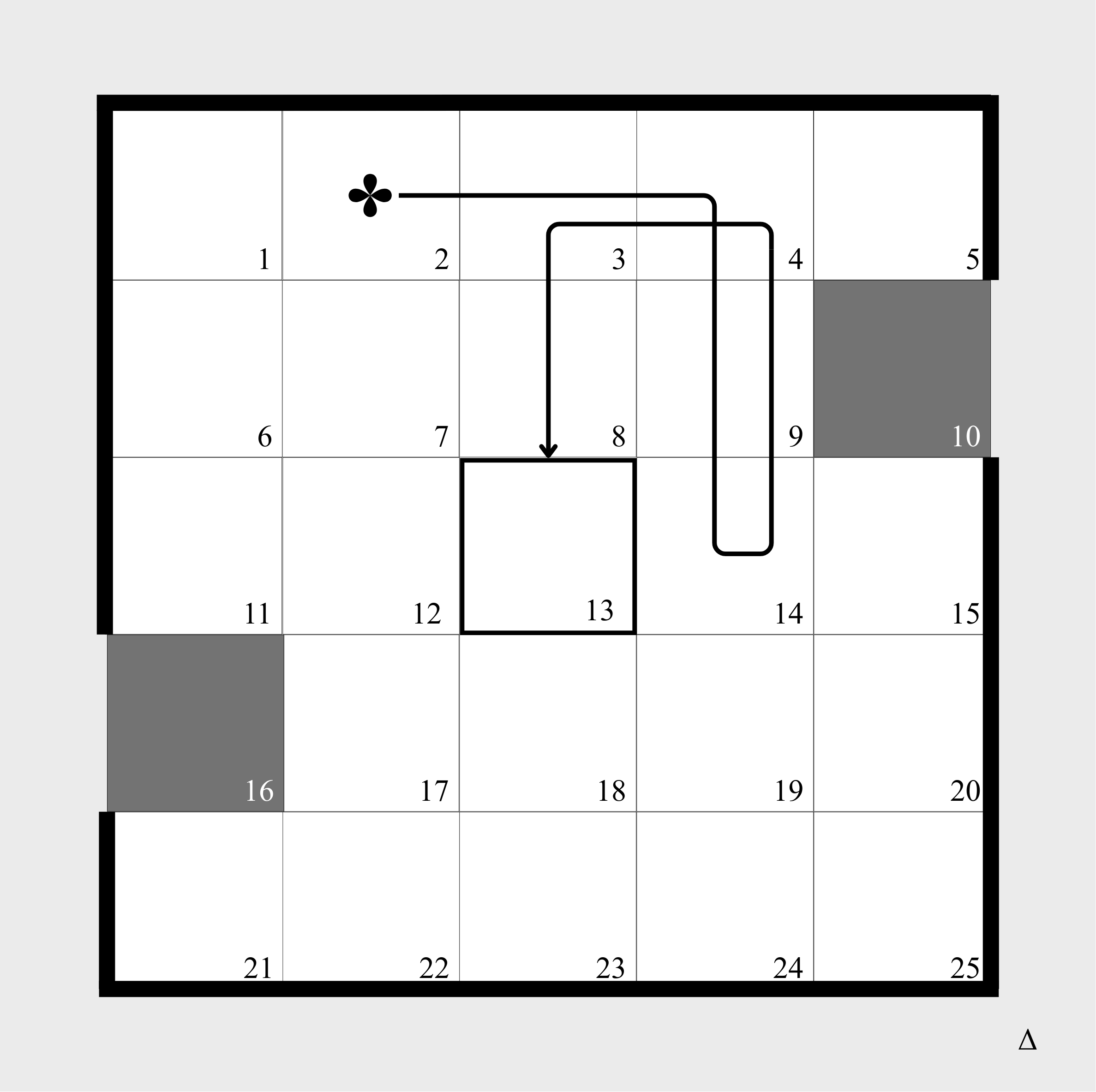}
        \caption{Drone in zone $2$ and initial state estimate = $\{1,2\}$.} 
        \label{fig:S1Z12D2}
    \end{subfigure}
    \hfill
    \begin{subfigure}{0.23\textwidth}
        \includegraphics[width=\textwidth]{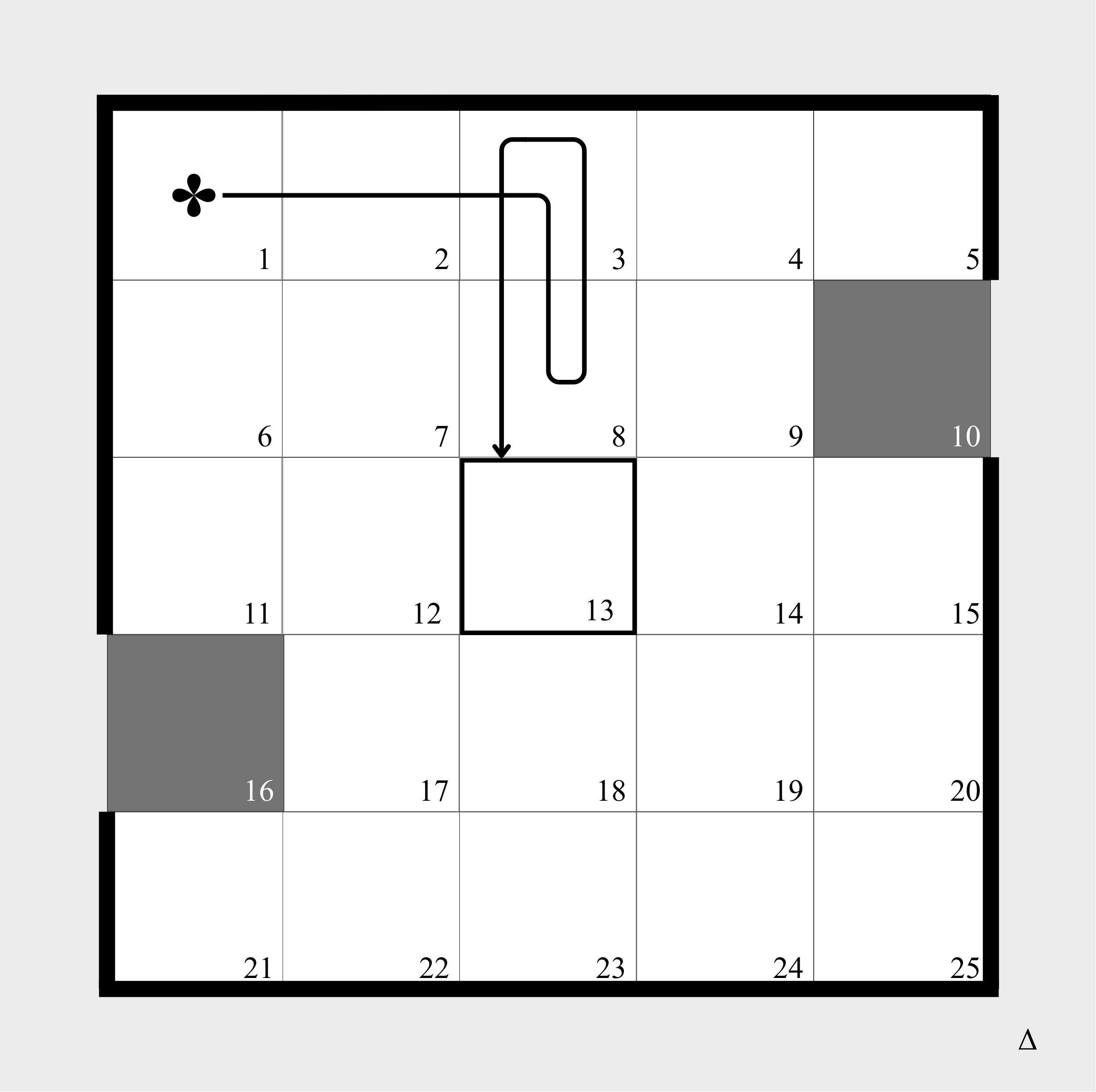}
        \caption{Drone in zone $1$ and initial state estimate = $\{1,2,6,7\}$.}
        \label{fig:S1Z1267D1}
    \end{subfigure}
    \hfill
    \begin{subfigure}{0.23\textwidth}
        \includegraphics[width=\textwidth]{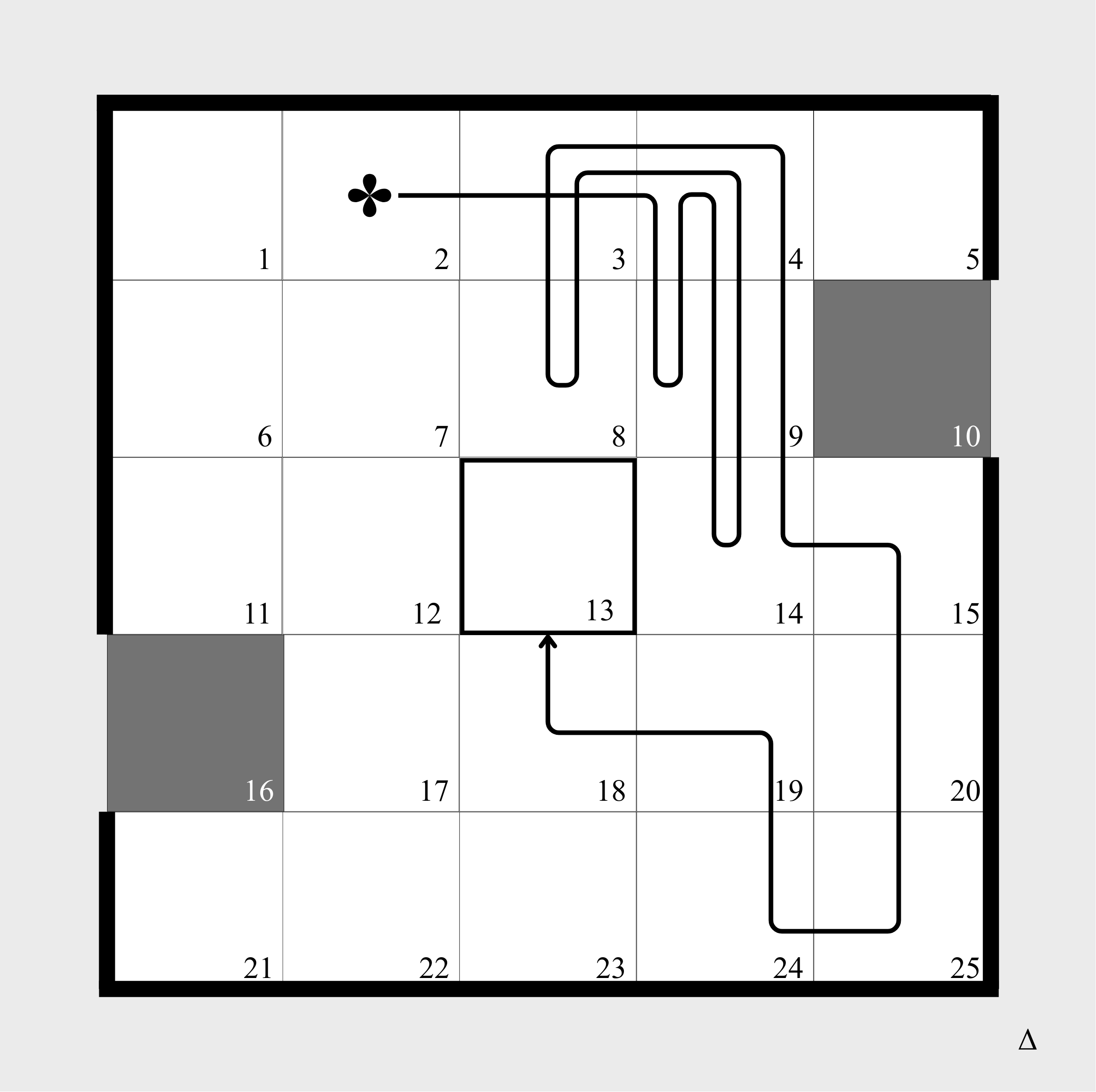}
        \caption{Drone in zone $2$ and initial state estimate = $\{1,2,6,7\}$.}
        \label{fig:S1Z1267D2}
    \end{subfigure}
    \caption{Recovery trajectories under different initial state estimates. The black quatrefoil represents the lost drone's initial position, and the arrow is the recovery path taken.}
    \label{fig:SimS1}
    \vspace*{-5mm}
\end{figure}

Figure~\ref{fig:SimS1} illustrates some of the recovery paths obtained in the simulations.
The examples highlight how different initial state estimates influence the resulting trajectories, even when the drone begins in the same physical zone.
In particular, the primary recovery supervisors synthesized from the estimates $\{1,2\}$ and $\{1,2,6,7\}$ lead to distinct maneuvering behaviors as the drone searches for a safe return path to the swarm.
Considering the drone is located in zone $1$, for the estimate $\{1,2\}$, the supervisor enables the event sequence: $m_e \rightarrow m_e \rightarrow m_s \rightarrow m_s$, a straightforward path.
However, for the state estimate $\{1,2,6,7\}$, the supervisor enables the event sequence: $m_e \rightarrow m_e \rightarrow m_s \rightarrow m_n \rightarrow m_s \rightarrow m_s$, passing twice through zones $3$ and $8$.
This behavior accounts for the possibility that the drone might be initially located in zone $6$ and it would have reached zone $13$ when the first $m_s$ event was enabled, consequently, the state estimate reached after the subsequent $m_n$ transition is considered unvisited and admissible during the RBTS construction, and thus this control action is enabled by the supervisor.
Furthermore, the algorithm from \cite{Cavalcanti:2025} can be adapted to use different selection criteria for picking admissible control actions (such as randomized choice or maximally permissive decisions) or different exploration methods (such as breadth-first or depth-first search), to obtain alternative recovery supervisors.

\begin{figure}[h]
    \centering
    \begin{subfigure}{0.1\textwidth}
        \includegraphics[width=\textwidth]{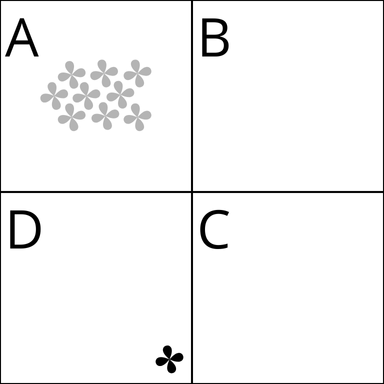}
        \caption{$t = 0s$.}
        \label{fig:SSA}
    \end{subfigure}
    \hfill
    \begin{subfigure}{0.1\textwidth}
        \includegraphics[width=\textwidth]{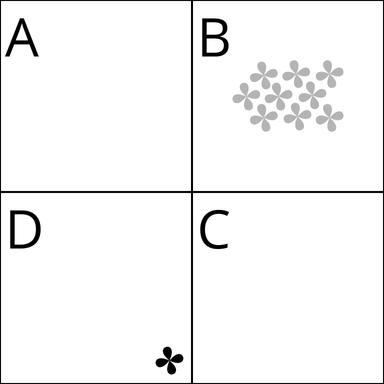}
        \caption{$t = 5s$.}
        \label{fig:SSB}
    \end{subfigure}
    \hfill
    \begin{subfigure}{0.1\textwidth}
        \includegraphics[width=\textwidth]{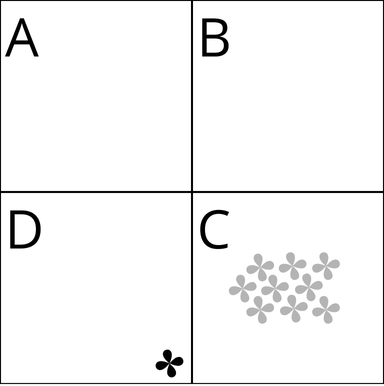}
        \caption{$t = 16s$.}
        \label{fig:SSC}
    \end{subfigure}
    \hfill
    \begin{subfigure}{0.1\textwidth}
        \includegraphics[width=\textwidth]{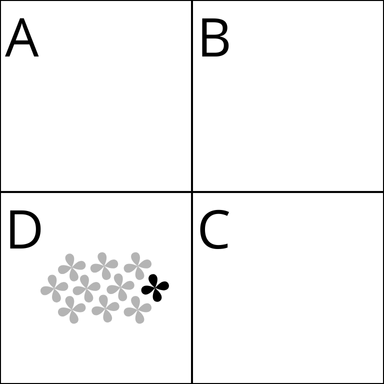}
        \caption{$t = 20s$.}
        \label{fig:SSD}
    \end{subfigure}
    \caption{Time-indexed snapshots of the secondary recovery strategy, depicting the drone’s return to the operational region and reconnection with the swarm.} 
    \label{fig:SimS3}
    \vspace*{-7mm}
\end{figure}

Additionally, the secondary recovery time, defined as the interval between the drone’s re-entry into the OR and its subsequent reintegration with the swarm, is also recorded. 
This strategy is illustrated in Figure~\ref{fig:SimS3}, with the corresponding simulation times shown in the subfigure captions.

\section{CONCLUSIONS}


 We address the problem of regrouping a drone with its swarm after it becomes lost due to faulty behavior or an attack.
We formalize this task within  SCT framework by interpreting the unintended separation of a drone from the swarm as a desynchronization between a plant and its supervisor.
Under this formulation, we apply the recovery strategy proposed in \cite{Cavalcanti:2025} to restore synchronization and reunite the swarm.
To do so, we model the mission map, the swarm’s exploration behavior, and the directional scanning logic activated when a drone detects its separation.
Simulated results show that the recoverability method can be implemented in this scenario, avoiding the reaching of unsafe states or regions out of range.
 However, the complexity of constructing an RBTS grows exponentially with respect to both the number of plant states and the number of events \cite{Cavalcanti:2025}. 
As a result, its online computation on real hardware remains a subject for future work.
One future direction we are exploring is new architectures, such as modular or decentralized ones, that can better treat different scenarios of detachment and regrouping, while considering local specifications for different drones and regions.
We are also currently working on relaxing some conditions related to recoverability in the DES framework to deal with the situation where the supervisor is not sure if the drone is in an unsafe region. 
Beyond this, future work will investigate additional SCT-based swarm implementations with the aim of more fully exploiting formal guarantees.

\addtolength{\textheight}{-12cm}   





\bibliographystyle{IEEEtran}
\bibliography{references}
\end{document}